\documentstyle[12pt,epsfig]{article}

\newcommand{\be}{\begin{equation}}
\newcommand{\ee}{\end{equation}}
\begin{document}  
\topmargin 0pt
\oddsidemargin=-0.4truecm
\evensidemargin=-0.4truecm
\renewcommand{\thefootnote}{\fnsymbol{footnote}}
\newpage
\setcounter{page}{0}
\begin{titlepage}   
\vspace*{-2.0cm}  
\begin{flushright}
FISIST/01-01/CFIF \\
hep-ph/0101116
\end{flushright}
\vspace*{0.5cm}
\begin{center}
{\Large \bf Solar Neutrinos with Magnetic Moment and Solar Field Profiles} 
\vspace{1.0cm}

{\large 
Jo\~{a}o Pulido
\footnote{E-mail: pulido@beta.ist.utl.pt}}\\  
{\em Centro de F\'\i sica das Interac\c c\~oes Fundamentais (CFIF)} \\
{\em Departamento de Fisica, Instituto Superior T\'ecnico }\\
{\em Av. Rovisco Pais, P-1049-001 Lisboa, Portugal}\\
\end{center}
\vglue 0.8truecm

\begin{abstract}
A statistical analysis of the pre-SNO solar neutrino data is made assuming 
the solar neutrino deficit to be resolved by the interaction of the neutrino 
magnetic moment with the solar magnetic field. Given the general characteristics
of the field profiles known to lead to the best event rate predictions, several 
specific choices of profiles are assumed and fits to the rates, to the electron 
recoil spectrum in SuperKamiokande and to the combined rates and spectrum are 
performed. Both the rate and spectrum fits are 
remarkably better than in a previous analysis of the magnetic moment solution 
which used a former standard solar model and previous data, and in particular, 
rate fits are far better than those from oscillations. The global analysis also
reflects the general very good quality of the fits. 
\end{abstract}
\end{titlepage}   


Whereas the apparent anticorrelation of the neutrino event rate with sunspot 
activity claimed long ago by the Homestake collaboration \cite{Homestake} 
remained unconfirmed by other experiments \cite{SuperK}, \cite{SAGE}, 
\cite{Gallex} and theoretical analyses \cite{Walther}, the magnetic moment 
solution to the solar neutrino problem is at present an important possibility 
to be explored in the quest for an explanation of the solar neutrino 
deficit. This is the idea, originally proposed by Cisneros \cite{Cisneros},
and later revived by Voloshin, Vysotsky and Okun \cite{VVO},  
that a large magnetic moment of the neutrino may interact with the magnetic 
field of the sun, converting weakly active to sterile neutrinos.
It now appears in fact that this deficit is energy dependent, in the sense
that neutrinos of different energies are suppressed differently. In order to 
provide an energy dependent deficit, the conversion mechanism from active to 
nonactive neutrinos must be resonant, with the location of the critical density 
being determined by the neutrino energy. Thus was developed the idea of the
resonant spin flavour precession (RSFP) proposed in 1988 \cite{LMA}. It involves
the simultaneous flip of both chirality and flavour consisting basically in the 
assumption that the neutrino conversion due to
magnetic moment and magnetic field takes place through a resonance inside matter 
in much the same way as matter oscillations \cite{MSW}. 
A sunspot activity related event rate 
in a particular experiment would hence imply that most of the neutrinos with energies 
relevant to that experiment have their resonances in the sunspot range. However,
the depth of sunspots is unknown (they may not extend deeper than a few hundred
kilometers) and the observed field intensity is too small in sunspots to allow
for a significant conversion. The anticorrelation argument has therefore lost 
its appeal for several years now.

Despite the absence of the anticorrelation argument,
several main reasons may be invoked to motivate RSFP and investigating its consequences
for solar neutrinos. In fact both RSFP and all oscillation scenarios indicate a 
drop in the survival probability from the low energy (pp) to the 
intermediate energy neutrino sector ($^7$Be, CNO, pep) and a subsequent moderate rise as 
the energy increases further into the $^8$B sector. The magnetic field profiles
providing good fits to the event rates from solar neutrino experiments typically show 
the characteristic of a sharp rise in intensity at some point in the solar 
interior, followed by a progressive moderate decrease \cite{PA}, \cite{Valle}. 
This is in opposite correspondence with the energy dependence of the probability  
in the sense that the strongest field intensities correspond to the smallest survival 
probabilities. Hence RSFP offers 
a unique explanation for the general shape of the probability, which naturally appears
as a consequence of the field profile. On the other hand, from solar physics and 
helioseismology such a sharp rise and peak field intensity is expected to occur along 
the tachoclyne, the region extending from the upper radiative zone to the lower 
convective zone, where the gradient of the angular velocity of solar matter is 
different from zero \cite{Parker}, \cite{ACT}. Furthermore, it has become 
clear \cite{GN}, \cite{PA}, \cite{Valle} that RSFP provides event rate fits 
from the solar neutrino experiments that are far better than all oscillation ones
\cite{BKS}, \cite{GG1}. Finally, there are recent claims in the literature for
evidence of a neutrino flux histogram \cite{SturrockS} containing two peaks, an
indication of variability pointing towards a nonzero magnetic moment of the neutrino.    

The aim of this paper is to present a statistical analysis of seven field profiles 
proposed in \cite{PA}, all obeying the general features described above, in the light 
of the updated standard solar model (BP 2000) \cite{BP2000} and the most recent data 
\cite{Homestake,SuperK,SAGE,Gallex}. For each profile the
best fits of both the rate and recoil electron spectrum in neutrino electron scattering
are determined along with the corresponding $\chi^2$. Unlike a previous analysis,
the contribution of hep neutrinos is taken into account in the spectrum fits.   
The results show a remarkable improvement in the quality of both sets of fits with respect 
to ref. \cite{PA} where the BP'98 standard solar model \cite{BP98} and the available data 
at the time were used. Throughout the calculations a value of $\mu_{\nu}=10^{-11}\mu_{B}$ 
is assumed. 

The profiles investigated are the following:

$Profiles~(1),(2)$
\be
B=0~~~,~~~x<x_R
\ee
\be
B=B_0\frac{x-x_R}{x_C-x_R}~,~x_{R}\leq x\leq x_{C}
\ee
\be
B=B_0\left[1-\frac{x-x_C}{1-x_C}\right]~ ,~x_{C}\leq x \leq 1
\ee
where $x$ is the fraction of the solar radius and units are in Gauss. For profile 1
$x_R=0.70$, $x_C=0.85$ and for profile (2) $x_R=0.65$, $x_C=0.80$. 

$Profile~(3)$
\be
B=0~~~,~~~x<x_R
\ee
\be
B=B_0\frac{x-x_R}{x_C-x_R}~,~x_{R}\leq x\leq x_{C}
\ee
\be
B=B_0\left[1-\left(\frac{x-0.7}{0.3}\right)^2\right]~,~x_{C}<x\leq 1
\ee
 
with $x_R=0.65$, $x_C=0.75$.

$Profile~(4)$
\be
B=0~~~,~~~x<x_{R}
\ee
\be
B=B_0\left[1-\left(\frac{x-0.7}{0.3}\right)^2\right]~,~x\geq x_{R},
\ee
with $x_R=0.71$.

$Profile~(5)$
\be
B=0~~~,~~~x<x_{R}
\ee
\be
B=\frac{B_0}{\cosh30(x-x_R)}~,x\geq x_{R}
\ee
with $x_R=0.71$.

$Profile~(6)$
\be
B=2.16\times10^3~~,~~x\leq 0.7105
\ee
\be
B=B_{1}\left[1-\left(\frac{x-0.75}{0.04}\right)^2\right]~,~0.7105<x<0.7483
\ee
\be
B=\frac{B_{0}}{\cosh30(x-0.7483)}~,~0.7483\leq x\leq 1
\ee
with $B_0=0.998B_1$.

$Profile~(7)$

\be
B=2.16\times10^3~~,~~x\leq 0.7105
\ee
\be
B=B_{0}\left[1-\left(\frac{x-0.75}{0.04}\right)^2\right]~,~0.7105<x<0.7483
\ee 
\be
B=1.1494B_{0}[1-3.4412(x-0.71)]~,~0.7483\leq x\leq 1.
\ee

The ratios of the RSFP to the SSM event rates $R^{th}_{Ga,Cl,SK}$ are defined as before 
\cite{PA} with the exception of the SuperKamiokande one in which the energy resolution
function \cite{Fukuda} is now taken into account 
\be
R_{SK}^{th}=\frac{\sum_{i}\int_{0}^{\infty}dE_e\int_{{E^{'}_e}_m}^{{E^{'}_e}_M}dE{'}\!\!_e
f(E{'}_e,E_e)\int_{E_m}^{E_M}dE\phi_{i}(E)[P(E)\frac{d\sigma_{W}}{dT^{'}}+(1-P(E))
\frac{d\sigma_{\bar{W}}}{dT{'}}]}
{\sum_{i}\int_{0}^{\infty}dE_e\int_{{E^{'}_e}_m}^{{E^{'}_e}_M}dE{'}\!\!_e
f(E{'}_e,E_e)\int_{E_m}^{E_M}dE\phi_{i}(E)\frac{d\sigma_{W}}{dT^{'}}}
\ee
Here $\phi_{i}(E)$ is the neutrino flux for component i ($i=hep,~^8\rm{B}$) and
$f(E^{'}_e,E_e)$ is the energy resolution function of the detector in terms of the
physical ($E^{'}_e$) and the measured ($E_e$) electron energy ($E_e=T+m_e$). The
lower limit of $E^{'}_{e}$ is the detector threshold energy (${E^{'}_e}_{m}={E_e}_{th}$ 
with ${E_e}_{th}$= 5.5MeV) and the upper limit is evaluated from the maximum 
neutrino energy $E_M$ \cite{PA}
\be
T^{'}_{M}=\frac{2E^{2}_{M}}{m_e+2E_{M}}.
\ee 
For the lower \cite {PA} and upper \cite{Homepage} integration limits of the neutrino 
energy one has respectively
\be
E_m=\frac{T^{'}+\sqrt{T{^{'}}^{2}+2m_eT^{'}}}{2}~,~E_{M}=15MeV~(i=^{8}\!\!B)~,~E_{M}=18.8MeV~
(i=hep).
\ee
The weak differential cross sections appearing in equation (17) are given by 
\be
\frac{d\sigma_W}{dT}=\frac{{G_F}^2 m_e}{2\pi}[(g_V+g_A)^2+
(g_V-g_A)^2\left(1-\frac{T}{E}\right)^2-({g_V}^2-{g_A}^2)\frac{m_{e}T}{E^2}]
\ee
for $\nu_{e}e$ scattering, with $g_V=\frac{1}{2}+2sin^2\theta_{W}$, $g_A=\frac{1}{2}$. 
For $\bar\nu_{\mu} e$ and $\bar\nu_{\tau} e$ scattering,
\be
\frac{d\sigma_{\bar{W}}}{dT}=\frac{{G_F}^2 m_e}{2\pi}[(g_V-g_A)^2+
(g_V+g_A)^2\left(1-\frac{T}{E}\right)^2-({g_V}^2-{g_A}^2)\frac{m_e T}{E^2}]
\ee
with $g_V=-\frac{1}{2}+2sin^2\theta_{W}$, $g_A=-\frac{1}{2}$.

The recoil electron spectrum in SuperKamiokande is now defined as
\be
R_{SK}^{th}=\frac{\sum_{i}\int_{{E_e}_j}^{{E_e}_{j+1}}dE_e\int_{{E^{'}_e}_m}^
{{E^{'}_e}_M}dE{'}\!\!_e
f(E{'}_e,E_e)\int_{E_m}^{E_M}dE\phi_{i}(E)[P(E)\frac{d\sigma_{W}}{dT^{'}}+(1-P(E))
\frac{d\sigma_{\bar{W}}}{dT{'}}]}
{\sum_{i}\int_{{E_e}_j}^{{E_e}_{j+1}}dE_e\int_{{E^{'}_e}_m}^{{E^{'}_e}_M}dE{'}\!\!_e
f(E{'}_e,E_e)\int_{E_m}^{E_M}dE\phi_{i}(E)\frac{d\sigma_{W}}{dT^{'}}}
\ee
with $i=hep,~^8\rm{B}$ for 18 energy bins (j=1,...,18) \cite{SuperK}. 

The fluxes and partial event rates for each neutrino component in each experiment were 
taken from \cite{BP2000} and the solar neutrino spectra from Bahcall's homepage 
\cite{Homepage}. The contribution of the hep flux to both the Gallium and Homestake 
event rates was neglected. The $\chi^2$ analysis for the ratios of event rates and 
electron spectrum in SuperKamiokande was done following the standard procedure 
described in \cite{PA} \footnote{Here only the main definitions and differences are 
registered. For the calculational details we refer the reader to ref.\cite{PA}.}.

The ratios of event rates to the SSM event rates and the recoil electron spectrum 
normalized to the SSM one, both denoted by $R^{th}$ in the following, were 
calculated in the parameter
ranges $\Delta m^2_{21}=(4-22)\times10^{-9}eV^2$, $B_0=(3-30)\times10^{4}G$ for
all magnetic field profiles and inserted in the $\chi^2$ definitions for the
rates and spectrum, 
\be
\chi^2_{rates}=\sum_{j_{1},j_{2}=1}^{3}({R}^{th}_{j_{1}}-{R_{j_{1}}}^{exp})\left[\sigma^{2}_{rates}
(tot)\right]^{-1}_{j_{1}j_{2}}({R}^{th}_{j_{2}}-{R_{j_{2}}}^{exp})
\ee
with (Ga=1, Cl=2, SK=3)
\be
\chi^2_{sp}=\sum_{j_{1},j_{2}=1}^{18}({R}^{th}_{j_1}-R^{exp}_{j_1})[\sigma^{2}_{sp}(tot)]^{-1}
_{j_{1}j_{2}}({R}^{th}_{j_2}-R^{exp}_{j_2})\,.
\ee

The quantities $R^{exp}$ in eqs. (21), (22) are directly read from tables I, II respectively
and the total error matrices $\sigma^{2}(tot)$ are derived from the definitions given in
\cite{PA}, using \cite{Homepage} and the error bars in tables I, II. 
In fitting the rates one has 3 experiments and 2 parameters (the mass square difference
between neutrino flavours and the value of the field at the peak), hence the number of 
degrees of freedom is one, while for the spectrum there are 18 data points, hence 16 
degrees of freedom. For global fits one has
\be
\chi^2_{gl}=\sum_{j_{1},j_{2}=1}^{21}({R}^{th}_{j_1}-R^{exp}_{j_1})[\sigma^{2}_{gl}(tot)]^{-1}
_{j_{1}j_{2}}({R}^{th}_{j_2}-R^{exp}_{j_2})\,.
\ee
Here index values 1, 2, 3 are used for the rates and 4, ..., 21 for the spectrum.  
The best fits for the rates in terms of $\Delta m^2_{21}, B_0$ and the corresponding values 
of $\chi^2$ for rates, spectrum and global fits are shown in table III.
In table IV the spectrum best fits are shown with the corresponding values of $\chi^2_{{sp}_{min}}$ 
and table V shows the global best fits with the values of $\chi^2_{rates}$, $\chi^2_{sp}$
in each profile. In this case the number of d.o.f. is 19 (=21-2). 

It is seen that the rate fits (table III) are in some cases excellent, with four of the 
profiles being much favoured, in particular profile 2. The most unfavoured ones (profiles 
4, 5 and 7) give nevertheless substantially smaller $\chi^2$ as compared to oscillations
\cite{BKS,GG1}. It is difficult at present to develop a general distinction between the
characteristics of those profiles which provide the best fits and the others. In profiles 
6 and 7 for instance the field intensity along the bottom of the convective zone rises in 
the same way and the quality of the rate fits is quite different: this suggests that
an upward facing concavity (profile 6) along the upper convective zone is preferred 
relative to a constant slope (profile 7).  In fact, since the low energy sector of the 
$^8B$ neutrinos is highly suppressed because their resonances are located close to the 
intermediate energy ones ($pep,CNO$) whose suppression has to be maximal, the higher energy 
sector of $^8B$ has to undergo little suppression in order to ensure a survival 
probability close to 1/2 as reported by SuperKamiokande. Hence the magnetic field 
should decrease steeply at first after its maximum and then more smoothly. The same 
conclusion is also
suggested from inspection of profiles 4 and 5: while they exhibit a sudden rise at the 
bottom of the convective zone with infinite slope, their downward slopes have opposite
shapes to each other, favouring the case with an upward facing concavity (profile 5).    
As for profile 4 the fits still worsen, as expected, if the exponent 2 in eq. (8) is
increased.
An infinite rise in field intensity (profiles 4 and 5) is also disfavoured relative 
to a finite one (compare 3 with 4 and 6 with 5). The case of the two triangle profiles
(1 and 2) suggests that the upward slope should be higher than the downward one.

Table III shows unusually large values of $\chi^2_{gl}$ at the rate best
fits for profiles 4 and 7. This is related to a high instability of $\chi^2$ both
for rates and global fits against small variations of $\Delta m^2_{21}$ and $B_0$
not shared by other profiles, as can be seen from a comparison with table V. 

The 90 and 99\% CL regions for the rates defined in the $\Delta m^2_{21},B_0$ plane 
from the condition $\chi^2\leq \chi^2_{min}+\Delta \chi^2$ with, for one degree of 
freedom, $\Delta \chi^2=2.71$ and 6.64 respectively, are shown in fig.1. The four 
profiles chosen are those for which $\chi^2_{min}<0.1$. 
As far as the spectrum fits are concerned, the results for the minima of $\chi^2_{sp}$ 
are also remarkably good (see table IV). A comparison
of tables III and IV and inspection of fig.1 show a discrepancy between the
rate and spectrum best fits. To further examine this discrepancy one can analise for
each profile the difference between the CL's corresponding to $\chi^2_{sp}$ at the rate and
spectrum best fits also listed in table IV \cite {RPP}. For 16 d.o.f. it is seen from
the values of this difference that all rate best fits lie within a region of 
10.5\% CL with respect to the spectrum best fit: the function $\chi^2_{sp}$ has a
weak dependence on the parameters $B_0$ and $\Delta m^2_{21}$, so that the mismatch 
between the rates and spectrum fits is statistically meaningless. Hence one 
should expect very good global fits. In fact the global best fits, all with a
remarkably low $\chi^2_{global}$ (see table V) 
always lie within a similar (11.4\% CL) region relative to the spectrum 
ones and also very close to the rate best fits, namely within a 
41\% CL region with respect to these, as can be derived from tables
III, IV and V and using ref. \cite{RPP}.

As an example, the predicted spectrum (eq. (22)) for profile 1 is shown in figs.2 and 3 
respectively at the rate and spectrum best fits superimposed on the SuperKamiokande data. 
The moderate rise occuring for $E_e\geq 12MeV$ is the effect of hep neutrinos.
 
As previously referred, a value of $\mu_{\nu}=10^{-11}\mu_{B}$ was assumed in the
calculations. Since the order parameter is the product $\mu B$ and the peak value of the 
field is probably as large as $3\times10^{5}G$ \cite{Parker,ACT}, the best fits imply an 
effective value of $\mu_{\nu}$ in the range $(2-5)\times10^{-12}\mu_{B}$ in consistency 
with most astrophysical bounds \cite{magmo}.

To summarize, the analysis of prospects for the magnetic moment solution to the solar
neutrino problem made on the basis of BP'2000 solar standard model and using the
most recent pre-SNO data reveals extremely good fits for the
rates for a class of field profiles with a steep rise across the bottom of the 
convective zone and a more moderate decrease up to the surface. There is a substantial 
and noteworthy improvement in the quality of the fits with respect to a previous analysis 
using BP'98 \cite{BP98} and previous data. Spectrum fits are also quite good and 
consistent with the rate fits.  

In the present analysis only time averaged data were considered and a fitting was made 
to a time constant profile 'buried' in the solar interior. If, on the contrary, the active
neutrino flux turns out to be time dependent, a situation most likely to be interpreted 
through the magnetic moment solution with a time dependent interior field, the present 
approach is obviously inadequate and all fittings thus made will in some way be flawed. 
Averaging the event rates over time implies disregarding possible information in the data 
which otherwise is available if different periods of time are considered \cite{SturrockS}. 
The robustness of such a procedure will greatly improve with the accumulation of more data.

\newpage


\begin{center}
\begin{tabular}{lcccc} \\ \hline \hline
Experiment &  Data      &   Theory   &   Data/Theory  &  Reference \\ \hline
Homestake  &  $2.56\pm0.16\pm0.15$ & $7.7\pm^{1.3}_{1.1}$ & $0.332\pm0.05$ & 
\cite{Homestake} \\
Ga     &  $74.7\pm5.13$ & $129\pm ^8_6$ & $0.59\pm0.06$ &
\cite{Gallex},\cite{SAGE} \\
SuperKamiokande&$2.4\pm0.085$ &
$5.15\pm^{1.0}_{0.7}$&$0.465\pm0.052$& \cite{SuperK}\\ \hline
\end{tabular}
\end{center}

{Table I - Data from the solar neutrino experiments. Units are SNU for
Homestake and Gallium and $10^{6}cm^{-2}s^{-1}$ for SuperKamiokande. The
result for Gallium is the combined one from SAGE and Gallex+GNO.}

\begin{center}
\begin{tabular}{cc} \\ \hline \hline
Energy bin (MeV)& $R_j^{exp}$ \\ \hline
$5.5<E_e<6$     & $0.461\pm0.025$ \\
$6<E_e<6.5$     & $0.438\pm0.02$  \\
$6.5<E_e<7$     & $0.455\pm0.016$ \\ 
$7<E_e<7.5$     & $0.474\pm0.016$ \\ 
$7.5<E_e<8$     & $0.490\pm0.018$ \\
$8<E_e<8.5$     & $0.477\pm0.018$ \\
$8.5<E_e<9$     & $0.455\pm0.018$ \\
$9<E_e<9.5$     & $0.463\pm0.021$ \\
$9.5<E_e<10$    & $0.472\pm0.026$ \\
$10<E_e<10.5$   & $0.460\pm0.027$ \\
$10.5<E_e<11$   & $0.452\pm0.027$ \\
$11<E_e<11.5$   & $0.473\pm0.033$ \\
$11.5<E_e<12$   & $0.457\pm0.037$ \\
$12<E_e<12.5$   & $0.429\pm0.041$ \\
$12.5<E_e<13$   & $0.488\pm0.049$ \\
$13<E_e<13.5$   & $0.493\pm0.058$ \\
$13.5<E_e<14$   & $0.583\pm0.062$ \\
$14<E_e<20$     & $0.505\pm0.078$ \\ \hline
\end{tabular}
\end{center}

{Table II - Recoil electron energy bins in SuperKamiokande (1117 days) and the corresponding
ratio of the experimental to the SSM event rate \cite{SuperK}.}

\begin{center}
\begin{tabular}{ccccc}\\ \hline \hline
Profile & Best fit ($\Delta m^2_{21},B_0$) & $\chi^2_{{rates}_{min}}$/1d.o.f. & $\chi^2_{sp}$/16d.o.f.
& $\chi^2_{gl}$/19d.o.f.\\ \hline
1        & ($7.08\times10^{-9}eV^2,~6.6\times10^4G)$ & $3.71\times10^{-2}$  &$9.85$  & $12.73$\\
2        & ($1.25\times10^{-8}eV^2,~1.28\times10^5G)$& $9.13\times10^{-4}$  &$11.1$  & $12.82$\\
3        & ($1.26\times10^{-8}eV^2,~9.5\times10^4G)$ & $5.96\times10^{-2}$  &$9.72$  & $13.24$\\ 
4        & ($1.7\times10^{-8}eV^2,~1.7\times10^5G)$  & $0.357$              &$11.45$ & $41.46$\\
5        & ($2.1\times10^{-8}eV^2,~1.42\times10^5G)$ & $0.133$              &$11.7$  & $11.93$\\
6        & ($1.65\times10^{-8}eV^2,~9.7\times10^4G)$ & $9.51\times10^{-3}$  &$11.35$ & $14.19$\\
7        & ($1.50\times10^{-8}eV^2,~2.3\times10^5G)$ & $0.38$               &$10.94$ & $81.44$\\ \hline
\end{tabular}
\end{center}  

{Table III - Rate fits: the values of $\Delta m^2_{21}$ and $B_0$ at the best fit 
of the rate for each profile, the value of $\chi^2_{{rates}_{min}}$ and the corresponding
values of $\chi^2_{sp}$ and $\chi^2_{tot}$.}
 
%

\begin{center}
\begin{tabular}{cccc}\\ \hline \hline
Profile  &  Best fit ($\Delta m^2_{21},B_0$)         & $\chi^2_{{sp}_{min}}$/16d.o.f. &
$CL(\chi^2_{{sp}_{min}})\!\!-CL(\chi^2_{{sp}_{rates}})$\\ \hline
1        & ($3.7\times10^{-9}eV^2,~4.3\times10^4G)$  & $9.46$ & $1.89\%$ \\
2        & ($6.4\times10^{-9}eV^2,~1.26\times10^5G)$ & $9.93$ & $6.70\%$ \\
3        & ($7.7\times10^{-9}eV^2,~9.5\times10^4G)$  & $9.48$ & $1.15\%$ \\ 
4        & ($\leq 10^{-11}eV^2,~8.66\times10^4G)$    & $9.74$ & $9.90\%$ \\
5        & ($9\times10^{-9}eV^2,~1.43\times10^5G)$  &  $9.94$ & $10.5\%$ \\
6        & ($7.8\times10^{-9}eV^2,~9.5\times10^4G)$ &  $10.0$ & $7.93\%$ \\
7        & ($\leq10^{-11}eV^2,~2.27\times10^5G)$    &  $9.75$ & $6.62\%$ \\ \hline
\end{tabular}
\end{center}  

{Table IV - Spectrum fits: the values of $\Delta m^2_{21}$ and $B_0$ at the best fit 
of the spectrum for each profile, the corresponding value of $\chi^2_{{sp}_{min}}$ and the 
difference between the confidence levels (or goodness of fits) of the spectrum and the  
rate best fits for 16 d.o.f.(see the main text for details). For profiles 4 and 7 
$\chi^2_{{sp}_{min}}$ is an asymptotic value for small $\Delta m^2_{21}$.}  

\begin{center}
\begin{tabular}{ccccc}\\ \hline \hline
Profile  &  Best fit ($\Delta m^2_{21},B_0$)         & $\chi^2_{{gl}_{min}}$/19d.o.f. &
$\chi^2_{rates}$/1d.o.f. & $\chi^2_{sp}$/16d.o.f.\\ \hline
1    & $(7.1\times10^{-9}eV^2,~6.7\times10^4G)$ & $9.94$ & $0.100$ & $9.79$  \\
2    & $(1.32\times10^{-8}eV^2,~1.27\times10^5G)$ & $11.83$ & $0.290$ & $11.33$  \\
3    & $(1.25\times10^{-8}eV^2,~9.4\times10^4G)$ & $9.81$ & $0.116$ & $9.63$  \\
4    & $(1.67\times10^{-8}eV^2,~1.69\times10^5G)$ & $12.7$ & $0.708$ & $10.95$  \\
5    & $(2.13\times10^{-8}eV^2,~1.44\times10^5G)$ & $11.88$ & $0.133$ & $11.82$  \\
6    & $(1.78\times10^{-8}eV^2,~9.7\times10^4G)$ & $12.29$ & $0.366$ & $11.83$  \\
7    & $(1.45\times10^{-8}eV^2,~2.28\times10^5G)$ & $12.0$ & $0.764$ & $10.16$  \\ \hline
\end{tabular}
\end{center}  

{Table V - Global fits: the values of $\Delta m^2_{21}$ and $B_0$ at the global best fit 
for each profile, the corresponding values of $\chi^2_{rates}$ and $\chi^2_{spectrum}$.} 


\newpage

\centerline{\large Figure captions}

\noindent
Fig. 1.
90\% and 99\% CL regions with respect to rate best fits on the 
$\Delta m^2_{21}$, $B_0$ (peak field value) plane. 
Units are $eV^2$ (x-axis) and Gauss (y-axis). From left
to right and top to bottom: profiles 1, 2, 3, 6. Circles denote the rate best
fits and diamonds the spectrum best fits (for the corresponding values of
$\chi^2$ see tables III and IV). The apparent mismatch between spectrum and
rate best fits is statistically meaningless as can be seen from the last
column of Table IV and the global fits (Table V).
\noindent

Fig. 2.
The electron recoil spectrum prediction for profile 1 for the values of
$\Delta m^2_{21}$, $B_0$ corresponding to the rate best fit (table III) as 
a function of the total electron energy superimposed on the data set for 1117 days 
\cite{SuperK}. For this curve $\chi^2_{sp}=9.85$ (see table III). Units are in eV.
Notice the moderate rise in the theoretical curve for $E_e\geq12MeV$ from the
effect of hep neutrinos.

Fig. 3.
Same as fig.2 for the values of $\Delta m^2_{21}$, $B_0$ corresponding to the
spectrum best fit. For this curve $\chi^2_{sp}=9.46$ (see table IV).

\newpage

\begin{figure}
\mbox{\psfig{figure=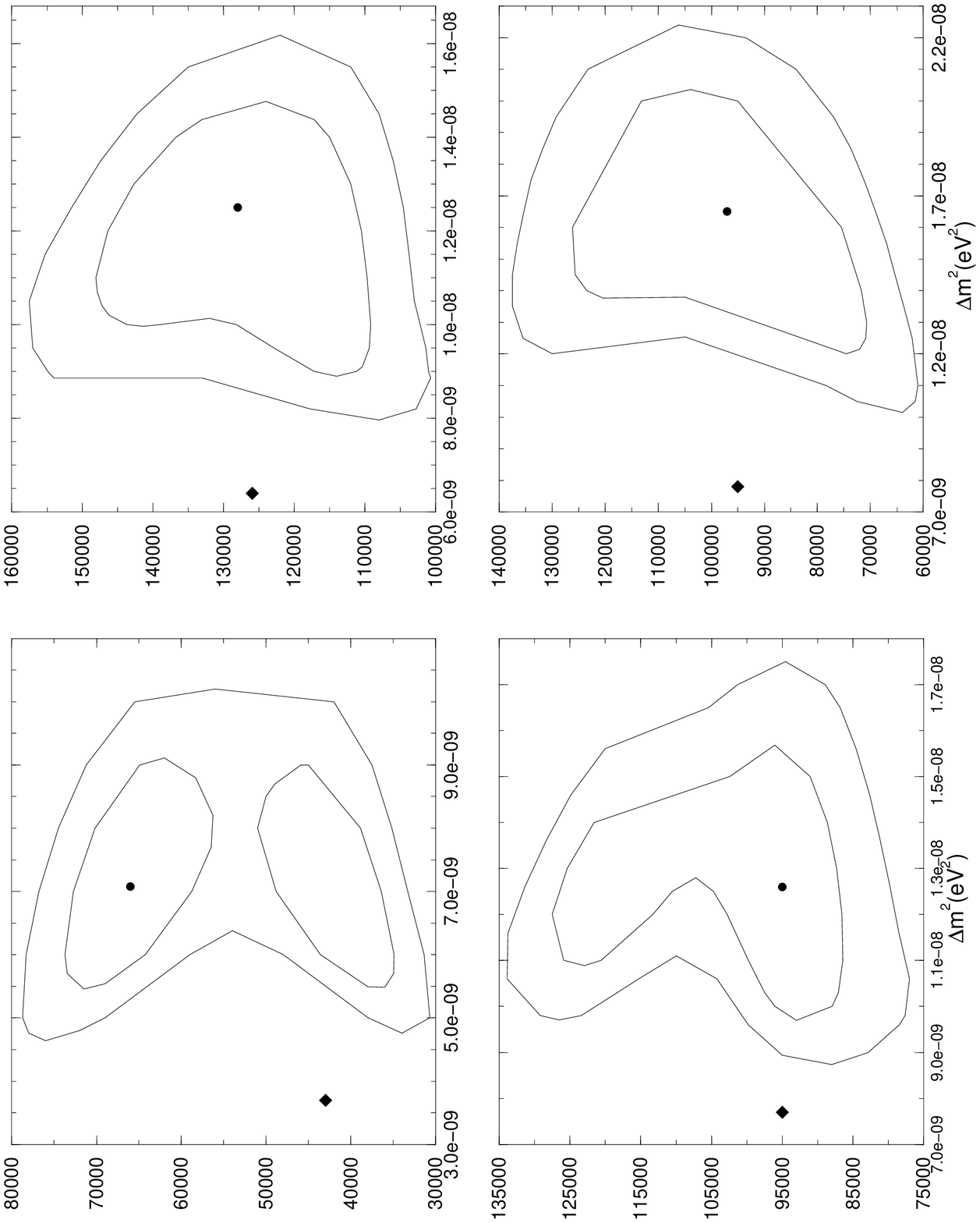,width=16cm}}
\centerline{\mbox{Fig. 1.}}
\end{figure}

\newpage

\begin{figure}
\mbox{\psfig{figure=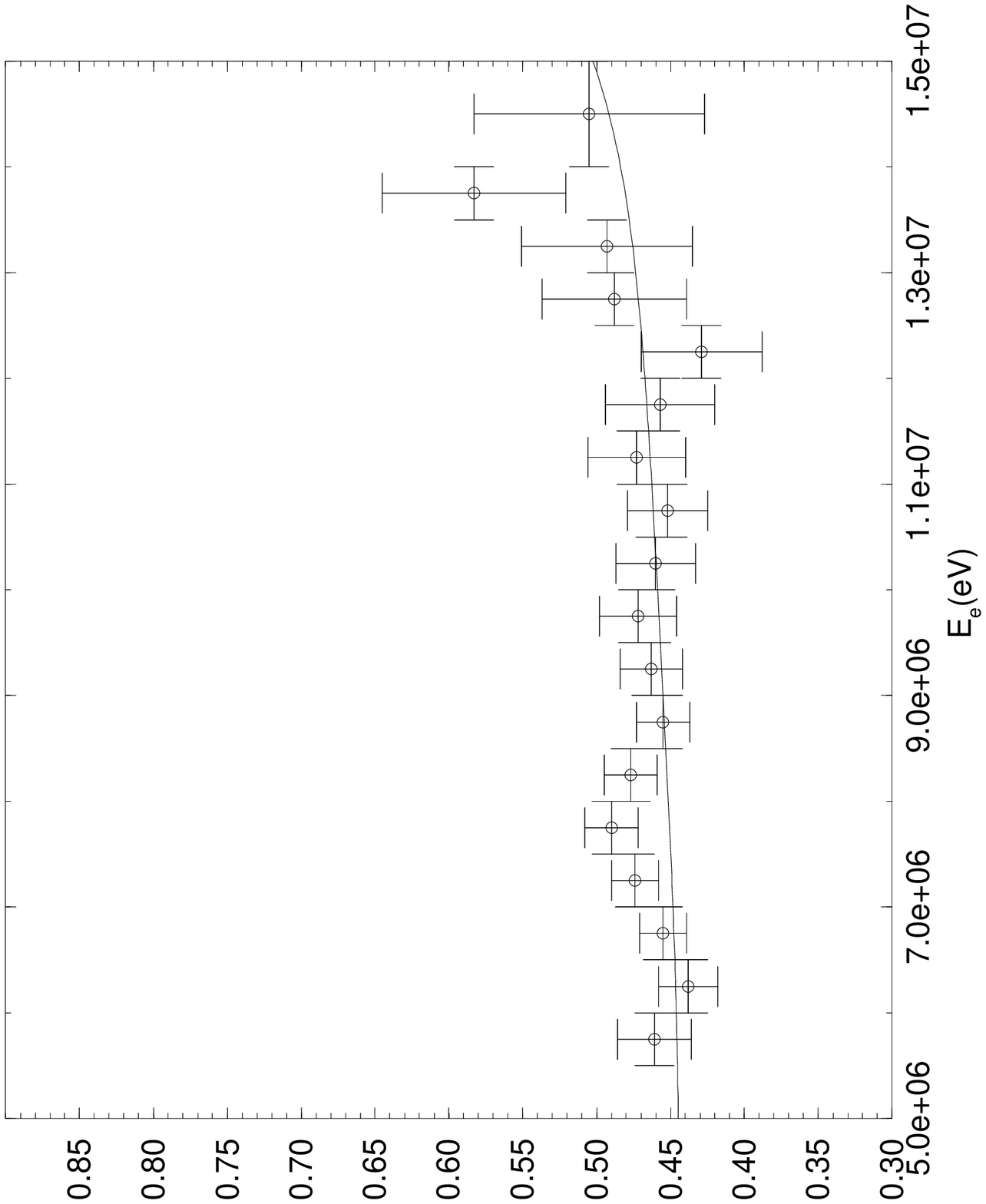,width=17cm}}
\centerline{\mbox{Fig. 2.}}
\end{figure}

\newpage

\begin{figure}
\mbox{\psfig{figure=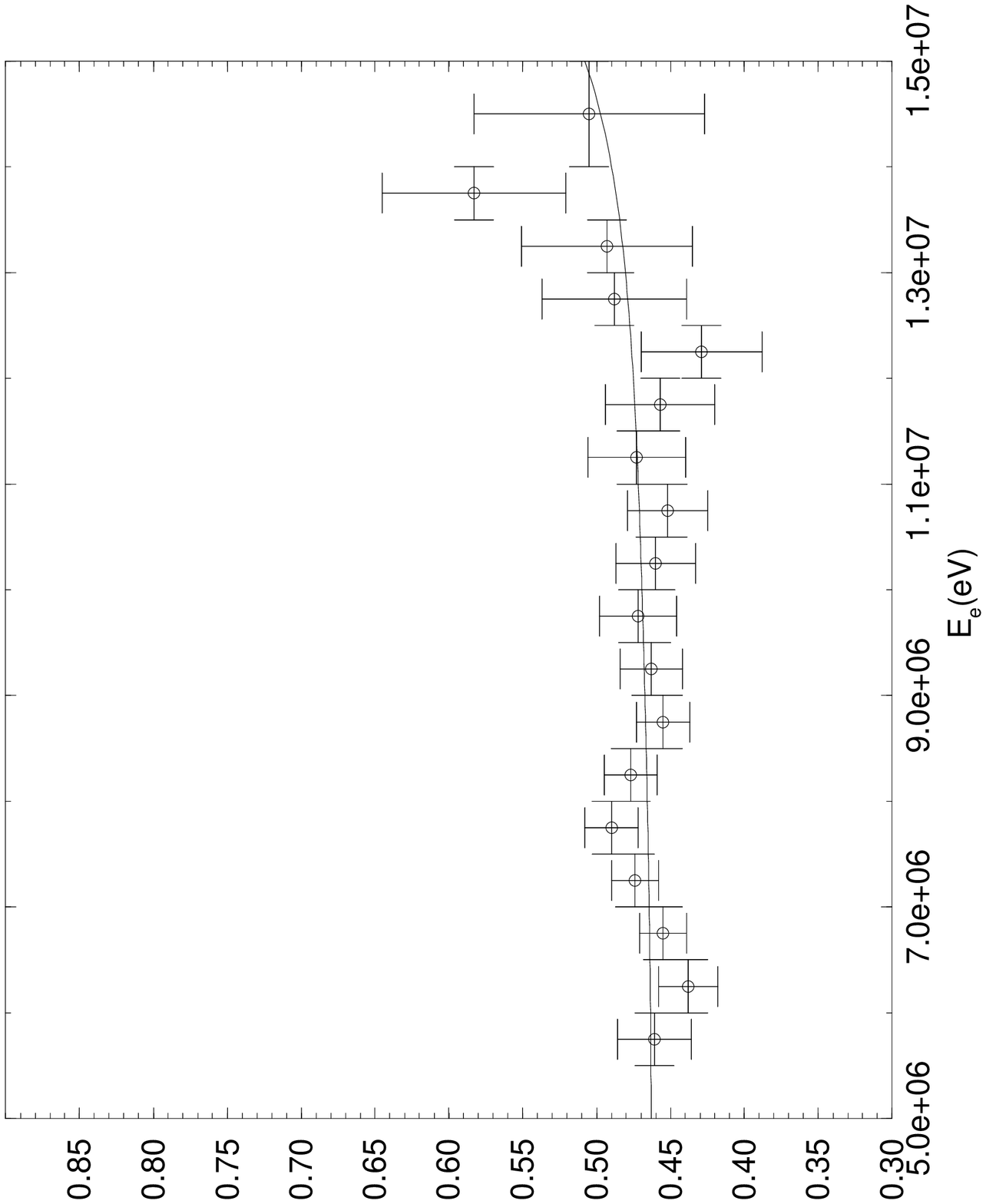,width=17cm}}
\centerline{\mbox{Fig. 3.}}
\end{figure}
\end{document}